\documentclass[a4paper,twoside]{article}

\usepackage{epsfig}
\usepackage{url}
\usepackage{caption}
\usepackage{subcaption}
\usepackage{calc}
\usepackage{amssymb}
\usepackage{amstext}
\usepackage{amsmath}
\usepackage{amsthm}
\usepackage{multicol}
\usepackage{pslatex}
\usepackage{apalike}
\usepackage{algorithm2e}
\usepackage[bottom]{footmisc}
\usepackage{SCITEPRESS}     % Please add other packages that you may need BEFORE the SCITEPRESS.sty package.

\usepackage{xcolor}

\begin{document}

\title{A Pragmatic Comparison of Cryptographic Computation Technologies for Machine Learning}

\author{%
\authorname{Marcus Taubert\sup{1}\orcidAuthor{0009-0002-3556-0262}, %
Adam Skuta\sup{1}\orcidAuthor{0009-0002-9892-7594} and %
Thomas Lorünser\sup{1}\orcidAuthor{0000-0002-1829-4882}}
\affiliation{\sup{1}AIT Austrian Institute of Technology, Vienna, Austria}
\email{Corresponding: Thomas.Loruenser@ait.ac.at}
}

\keywords{Secure Computation, Machine Learning, Fully-Homomorphic Encryption, Secure Multi-Party Computation}
%The paper must have at least one keyword. The text must be set to 9-point font size and without the use of bold or italic font style. For more than one keyword, please use a comma as a separator. Keywords must be titlecased.

\abstract{
As security demands increase, the importance of secure computation technologies grows, yet these technologies can often seem overwhelming to practitioners.
Furthermore, many approaches focus only on a single technology, potentially overlooking superior alternatives.
This work aims to address the issue of selecting the right technology for secure computation by presenting a comparative analysis of two highly relevant cryptographic methods and their software implementations, with a particular focus on machine learning.
Firstly, we provide a theoretical summary and comparison of the secure computation paradigms of secure multi-party computation (SMPC) and fully homomorphic encryption (FHE). We outline the advantages and limitations of the protocols, as well as the relevant open-source software implementations.
Secondly, we present the results of extensive benchmarking of the main software frameworks identified for machine learning operations and models.
Regarding the current state of the art in FHE, we observe that it outperforms SMPC for regressions. 
Additionally it may be faster for simple dense networks using GPUs or Hybrid Models.
Conversely, SMPC showed superior performance for complex models such as CNNs.
Our results should pave the way for more technology-agnostic benchmarking of secure computation technologies for machine learning, providing guidance for practitioners looking to adopt these technologies. 
}

%With ever increasing security demands secure computation technologies get more and more popular. At the same time they often seem overwhelming for practitioners foreign to these subjects. Additionally most of them focus on one technology while completely disregarding a possible better solution. We summarize the two secure computation technologies multi party computation~(MPC) and fully homomorphic computation~(FHE) and give an outline of their advantages and disadvantages. Furthermore we present benchmarks results over machine learning operations and models and find, that FHE works really well for regressions and simple dense networks. Subsequently, MPC should be used for more complex models like CNNs, LSTMs and Transformer.

\onecolumn \maketitle \normalsize \setcounter{footnote}{0} \vfill

\section{\uppercase{Introduction}}%
\label{sec:introduction}
%ICISSP
%\url{https://icissp.scitevents.org/}

In an increasingly digital world with more and more data flowing through the internet, privacy is an ever-increasing concern for users, governments, and organizations.
This has led to a growing need for fast, reliable, and secure cryptographic technologies, which is why we have seen so much research on the topic in the last decade.
Especially the ability to perform calculations on encrypted data gained a lot of traction in recent years. 
Fully-Homomorphic Encryption (FHE) and Secure Multi-Party Computation (SMPC) are such methods which allow for computations on encrypted data and therefore support secure outsourcing scenarios or enable collaboration among mutually distrusted stakeholders.
Moreover, during the last decade, theory has been transformed into concrete schemes and software frameworks that are now fast enough for application in practice.
Despite of FHE and SMPC having rather different security assumptions (non-collusion vs. LWE-related hardness), from a user's perspective both technologies can potentially be used to increase security in most scenarios, especially if the computation is done on third-party cloud providers.
SMPC naturally supports single and multiple input and output parties, while FHE natively support a single input/output party, but has also been adapted to multi-party use cases using threshold FHE, thus they provide similar functionalities.
Therefore, it is reasonable to compare the technologies for a given set of computing tasks in order to inform technology selection for particular use cases.
We will examine how each technology performs for different machine learning use cases.

\subsection{Motivation}

Our work is motivated by the fact that, despite using different internal techniques, SMPC and FHE can fulfil similar needs.
This makes it difficult for practitioners to understand and select the appropriate technology~\cite{BachlechnerHKLR25}.
Our goal is to address this issue by comparing and benchmarking the two technologies in various machine learning tasks, in order to determine their respective advantages and disadvantages. 
This will provide engineers with a clearer picture of their capabilities and expected performance in typical machine learning scenarios.
Ultimately, this work represents the first step in developing a technology-agnostic, problem-oriented secure computation benchmarking methodology.
This will facilitate the extraction of recommendations for selecting technologies in specific circumstances to achieve optimal efficiency and privacy, which, in turn, should encourage adoption and further integration \cite{FabianekKLS24}.

\subsection{Related work}

\paragraph{SMPC.} A lot of different SMPC frameworks have been in development in recent years. MP-SPDZ~\cite{keller_mp-spdz_2020} is a scientific framework designed for evaluating and benchmarking different programs with a wide variety of protocols supporting all adversary types.
It also supports importing deep learning models and allows their training in a privacy-preserving manner.
MPyC is a general SMPC framework, that supports a wide variety of computations written in Python. 
Recent advancements in privacy-preserving machine learning using SMPC have focused on achieving acceptable performance on Large Language Models~(LLMs)~\cite{dong_puma_2023,akimoto_privformer_2023,luo_centaur_2024}. 
These advancements use different approaches, such as simplifying activation functions to reduce inference time and knowledge distillation to achieve acceptable accuracy.
Alternatively hybrid models are used where the client evaluates the neural network's activation functions in plaintext or permutes the model's weights to keep them public, which speeds up inference time while preventing model stealing \cite{SkutaSHHL25}.

\paragraph{FHE.}
In recent years, there has been a lot of research regarding FHE benchmarks for basic operations.
The paper~\cite{fhe_benchmarks} tests basic operations, such as addition, multiplication and rotation over different FHE schemes.
"Cross-Platform Benchmarking of the FHE Libraries: Novel Insights into SEAL and OpenFHE" \cite{fhe_open_seal_benchmark} on the other hand compares two well-known libraries by benchmarking simple calculations and finds, that OpenFHE outperforms the SEAL library.
%Another paper examines calculation times for bootstrapping in various schemes \cite{bootstrap_time}.
%TFHE is here the fastest candidate by large margin. \\
There also have been made numerous advances in regards to machine learning with FHE. 
FHEMaLe \cite{FHEMaLe} is a framework designed to test inference times on the cloud with CKKS and an edge device using TFHE.
They implement the k-nearest-neighbour, support vector machine, logistic regression as well as a neural network.
CKKS performed better in all cases, with the k-nearest-neighbor model performing the worst for TFHE.
Orion \cite{Orion} is a fully automated framework for translating neural networks into FHE and running single-threaded inference.
The authors use the CKKS scheme and benchmark their framework on different neural networks such as various ResNet configurations and Dense Models.
They also propose and integrate a single-shot multiplexed packing strategy for arbitrary convolutions and an automatic bootstrap placement algorithm.
Ultimately, they present the results of encrypted image recognition evaluations.

\subsection{Contributions}

To the best of our knowledge, we provide the first pragmatic comparison of secure computation technologies based on cryptography, i.e.,  SMPC and FHE, with a focus on machine learning.
Therefore, we first summarize and compare both technologies on a theoretical, basis also considering major software implementations.
Secondly, we do a thorough performance benchmark on general purpose hardware for most relevant operations used in machine learning.
We also benchmark basic machine learning models to give concrete results of what is possible with currently available FHE and SMPC implementations.
Lastly, we summarize our findings and provide some guidance for practitioners.

The remainder of the paper is structured as follows.
We begin by summarizing and comparing the technologies used in this study in Section \ref{sec:protocol_comparison}.
An overview of the software frameworks employed can be found in Section \ref{sec:software_frameworks}, where we also highlight their main characteristics. 
Section \ref{sec:benchmarks} follows with a series of benchmarks covering essential ML operations as well as machine learning models.
The paper ends with a summary of the key insights and implications of our findings in Section \ref{sec:discussion_conclusion}.

\section{\uppercase{Protocol Comparison}}\label{sec:protocol_comparison}

In this section we summarize SMPC and FHE, before comparing them and extracting conceptual advantages and disadvantages.

\subsection{Secure Multi-party Computation}

Secure Multi-Party Computation~\cite{evans2018pragmatic} is a set of cryptographic protocols, that allows a group of parties to collaboratively evaluate a function over their respective private inputs, while ensuring that no participant learns any information about the others' data.
The objective is to obtain the same output that would results if all inputs were centrally computed on by a fully trusted third-party.
More formally, SMPC considers a setting with $n$ parties $P_1,P_2,...,P_n$, where each party $P_i$ possesses a private input $x_i$. The task is to compute a function~\ref{eq:smpc-general}, such that the output $y$ is revealed only to the designated party or parties, with no additional information leakage.
\begin{equation} \label{eq:smpc-general}
    y \leftarrow f(x_1,x_2,...,x_n)
\end{equation}

The role of SMPC protocols is to ensure the privacy and integrity of the underlying computation. These protocols are often based on the security model they support~(semi-honest, malicious and covert) and the number of adversary parties in the computation~(honest-majority, dishonest-majority).
% Based on the adversary types of the corrupted parties, the protocols can be categorised as follows:
% \begin{itemize}
%     \item \textbf{Semi-honest} adversaries~(also know as honest-but-curious) are corrupted parties that follow the computation correctly, but can attempt to learn some of the information, that should stay unknown to them. They can collude with other parties as well,
%     \item \textbf{Malicious} adversaries deviate from the protocol and may actively attempt to compromise the privacy and correctness of the computation,
%     \item \textbf{Covert} adversaries do not have to follow the protocol and can alternate between malicious and semi-honest behaviour, depending on their underlying strategy.
% \end{itemize}

% \noindent 
% The security model may alternatively be defined by the number of corrupted parties, which leads to two primary categories:
% \begin{itemize}
%     \item \textbf{Honest-majority} setting assumes, that more then half of the party members are honest and not corrupted,
%     \item \textbf{Dishonest-majority} setting assumes, that half or more of the party members are corrupted.
% \end{itemize}

Secret-sharing is a cryptographic technique, that serves as a fundamental component in SMPC systems. It divides a secret value into 
$n$ shares, which are distributed among $n$ parties such that only a predefined subset of parties is required to reconstruct the original secret. 
Once distributed, these shares can be used to perform computations, supporting operations such as addition and multiplication, although some operations may require additional communication between parties. 

\subsection{FHE Protocol}

Fully-homomorphic encryption often regarded as the holy grail of cryptography, enables performing calculations on encrypted data and therefore keeping information secret.
%It relies on the hardness assumption of the learning with errors problem which, from the current research standpoint, is said to be post quantum computing secure\cite{LWE_quantum}.
A common use case involves a client outsourcing computation to an untrusted cloud server.
Because of it's homomorphic/deterministic characteristics, FHE is not secure against active attacks and only fulfils IND-CPA or IND-CPA-D \cite{CPA-D} requirements.
Thus a general trust model of a semi-honest server needs to be established.
%which entails the server's objective of acquiring as much information as possible by observation only.

To prevent dictionary attacks, it is necessary for each ciphertext to contain some randomness called noise, which grows each operation.
At a certain point it grows beyond allocated free space and overwrites data.
This makes correct decryption impossible and limits circuit size. 
To allow for further operations a so called bootstrap is used, which resets the noise.
Historically this has been very slow \cite{bootstrap_time}.
Consequently two different approaches emerged.
Levelled schemes aim to adjust their parameters, such that the entirety of the circuit fits into the noise budget.
In contrast, bootstrapped schemes aspire to minimize calculation overhead from bootstrapping \cite{Joye2021HomomorphicEncryption101}.

Levelled schemes, such as BFV~\cite{BFV} and CKKS~\cite{CKKS}, are usually faster for basic linear calculations.
However their approach presents significant challenges fitting larger models in the noise budget.
Thus TFHE~\cite{TFHE}, a bootstrapped scheme, is used.
It's bootstrapping procedure takes milliseconds or less, which gives it an obvious advantage.
It operates over the Torus, thereby implementing boolean arithmetic.
Furthermore it implements programmable bootstrapping, a feature, that facilitates the evaluation of any univariate function for free during the bootstrap.
% This is achieved by creating look-up-tables, which map all possible inputs to outputs. 
% Subsequently, the ciphertext is employed to select the corresponding table entry, which is substituted for the original ciphertext.
% Concurrently, this refreshes the ciphertext and mitigates noise.
The combination of look-up-tables and boolean gates enables the execution of arbitrary functions.
Thus we use this scheme in the following benchmarks and whenever we talk about FHE, we implicitly mean the TFHE scheme.  

\subsection{Protocol Discussion and Comparison}

SMPC has a few notable advantages for privacy-preserving computation. Due to it's multi-party nature it offers decentralization, eliminating the single point of failure problem, and allows the computation cluster to be configured in a way, that tolerates failure/corruption of other parties. SMPC also supports a wide variety of protocols with varying trust models, that can be configured and tailored to the specific requirements of the application.

However SMPC also comes with several drawbacks. The most significant one is, that this approach requires $n$ times the amount of hardware for $n$ computing parties for an evaluation, that would usually require one party. This leads to an increase in deployment cost. Due to its decentralised nature, the computation is heavily affected by network artifacts, such as delay, packet loss and bandwidth, decreasing the added value of specialized hardware like GPUs.

Compared to SMPC, FHE is rather easy to deploy. There is only the need for one server entity doing all calculations and the clients handling encryption and decryption. This makes it quick and easy to set up and commission new FHE infrastructure. It is also possible to convert existing servers to FHE relatively simple, without having to build new systems or networks. 

On the other hand, as we will see later, the TFHE fully-homomorphic encryption scheme right now is quite slow even for simple arithmetic calculations, making it practically infeasible to use in real world scenarios. The reason for that is the needed continuos bootstrapping in order to reduce noise or evaluate functions like boolean gates. At the same time looking back, bootstrapping needed minutes to compute, and with some schemes still does \cite{bootstrap_time}. TFHE has been able to cut this down to only milliseconds \cite{TFHE,fast_bootstrap}. That means with more research and development we can expect even more speed-up. Furthermore it is possible to increase execution efficiency through the use of GPUs or FHE based hardware. According to Zama’s benchmarks, computations performed on GPUs were about ten times faster than those executed on CPUs. So while it may not be feasible now, rapid improvements are made and potential usability might be achieved in the near future. 

Another disadvantage is FHE's homomorphic nature. Currently there are no fully-homomorphic schemes, that fulfill any IND-CCA security notion \cite{CCA1_security}. This means FHE is only able to function securely for semi honest servers, which try to get as much information by observation only. However it might be possible achieving IND-CCA security using SNARKs as stated here \cite{vCCA_security}.

\section{\uppercase{Software Frameworks}} \label{sec:software_frameworks}

In this section we compare two software frameworks, that we use to implement all following benchmarks. For the FHE part we decided on using Concrete-ML. There are numerous other frameworks like OpenFHE \cite{OpenFHE} or SEAL \cite{seal}, but it is the only one that specializes in machine learning implementing various models. It additionally has good documentation and maintenance. For SMPC SPU is the only machine learning framework, that is currently not deprecated. Another viable framework, which is not a purely machine learning library but supports it, would have been MP-SPDZ \cite{keller_mp-spdz_2020}, but it is not nearly as user friendly as SPU.

\subsection{SMPC: Secretflow-SPU}

Secretflow-SPU~\cite{ma_secretflow-spu_2023} is a general-purpose framework for Privacy-preserving Machine Learning (ppML) designed to simplify the development of SMPC-based machine learning models for researchers. It comprises a frontend compiler, which exposes a Python API, and a backend runtime. The frontend accepts an ML program and compiles it into an intermediate representation known as Privacy-preserving High-Level Operations (pphlo). The backend subsequently interprets and executes the pphlo on a virtual device composed of multiple interconnected nodes that implement a configurable MPC protocol.

The compilation process into this intermediate representation is inspired by non-MPC ML compilers such as Google’s XLA. XLA employs High-Level Operations (HLO) as its intermediate representation, capturing the computational graph of the ML model before applying hardware-specific optimizations and lowering it to machine code for execution on CPUs, GPUs, or TPUs. SPU adopts a similar strategy: it utilizes XLA to generate the HLO representation, then applies additional compilation stages that annotate the computational graph with privacy semantics and data types, finally producing the pphlo representation.
% Secretflow-SPU is chosen as a framework of choice due to it's ML-centric nature and ease of development.

\subsection{FHE: Concrete-ML}

Concrete-ML is an open-source library operating under the BSD-3 license that is developed by the Zama Foundation \cite{ConcreteML}. It is a machine learning framework designed to simplify the implementation of regression and classifier models, as well as neural networks. It consists of a Python API frontend for an MLIR-based compiler and a backend runtime, backed by its own Concrete and TFHE-rs libraries \cite{concrete_arch}. It uses the TFHE fully-homomorphic encryption scheme under the hood. Developers can choose to use pre-implemented regression or classifier models, which are based on scikit-learn implementations, or they can insert their own Torch, ONNX or Brevitas-based neural networks. Concrete-ML offers API calls that automatically convert these into their FHE equivalents and run inference on them. To adhere to the TFHE's restrictions, the compiler scales all floating-point numbers to integers and then uses quantization to scale them down to eight bits or fewer. Quantization is necessary, because programmable bootstrapping is only usable with up to 16-bit integers. By adjusting the magnitude of quantization, developers can balance speed with smaller bit sizes and accuracy with larger ones.
% Furthermore Concrete-ML provides the ability to create servers and clients without specifying intricate distributed server logic, such as sockets and transmission protocols. In this regard, it also implements hybrid models. These models allow developers to split model inference between the client and server. By performing non-linear and FHE expensive calculations at client side and encrypted ones on the server, a considerable speed-up can be achieved while maintaining accuracy and confidentiality. \\

\subsection{Framework Comparison}

To compare the two frameworks, we selected a handful of criteria. To measure the size of the libraries we first examined the number of lines of code. We used the command-line interface tool cloc\cite{cloc}, which calculates the number of blank lines, comments and code lines for all files in a given repository, depending on their programming language. Then, we added together all the code lines of the relevant programming languages. The programming language tells us what the practitioners use and what the backend was written in. Next, we examined the level of documentation, maintenance, and number of supported models. We evaluate how well each criterion is met and assign one to three pluses to it, with three being the best and one the worst. For documentation, we consider the presence of documentation, examples, comments in the code and forums with user questions. In order to determine the level of maintenance, we examine the number of Git commits over time, the version history and open issues. For supported models, we compare the frameworks' capabilities and look at the types of models they integrate. The final "Ease of Use" row summarizes how approachable each framework is for developers and reflects the overall learning curve, API intuitiveness, and setup effort.   

\begin{center}
\begin{table}[h]
\centering
\caption{Comparison of both software libraries.}    
\begin{tabular}{ c|c|c } 
    \centering
    & SPU & Concrete-ML\\
    \hline
    lines of code & $\approx$150k& $\approx$600k\\
    prog. language & Python,C++ & Python, Rust \\
    Documentation & + & +++ \\ 
    Maintenance & + & ++ \\
    supported models & +++ & ++ \\
    easy of use & +++ & +++\\
\end{tabular}
\label{fig:framework-comparison}
\end{table}
\end{center}

As shown above Concrete-ML is a considerably larger framework. This is due in part to the fact, that it relies on the underlying Concrete and TFHE-rs libraries, which make up most of code. Both libraries offer a Python API interface and are written in C++ or Rust for fast and save execution. We regard Concrete-ML's documentation as very good because it includes examples, documentation, videos and an active community. SPU on the other hand, has good examples to get started, but suffers from minimal documentation of the public Python API, some of which is not translated to English. Similarly, SPU isn't that well maintained and receives few commits per month. New features are rare and most updates are basic maintenance, such as updating documentation, upgrading dependencies and fixing bugs. Concrete-ML on the other hand is well maintained, though we have noticed a gradual decline in commit frequency over the last few months. The same is true for the underlying library Concrete. However, TFHE-rs is very well serviced. Both libraries support a wide variety of models. In SPU developers can implement any model, as long as it is written in JAX, earning it three pluses. Concrete-ML doesn't support more complex models like LSTMs or Transformers. These models are very important, so Concrete-ML only receives two. SPU and Concrete-ML both receive the full three pluses in the last category. They are very easy to use because they both consist of a concise and self-explanatory Python API and developers don't need deep understanding of the secure computation technologies in order to use them.

\section{\uppercase{Benchmarking}}\label{sec:benchmarks}

In this section we compare these two cryptographically secure computation technologies with regard to calculations frequently used in machine learning and such models.
To do so, we test basic operations and subsequently evaluate regression and deep learning models.
All tests have been performed on a random sample, such that its standard deviation is less then one-tenth of its mean.

\subsection{Basic Operations}

This section compares the basic operations in the context of neural networks. Four operations were tested in this regard. The first one is matrix multiplication, a fundamental component of any neural network. We tested the inference speed of three different matrix sizes in both Secretflow-SPU and Concrete-ML. The other basic operations are three activation functions, which also belong to the basic building blocks of neural networks. The tested functions are Rectified Linear Unit~(ReLU), Gaussian Error Linear Unit~(GELU) and Sigmoid. 

\paragraph{Matrix Multiplication.} 
The results are displayed in Table~\ref{tab:matmul-smpc-fhe-time}. As can be observed Concrete-ML achieves better results for small matrices, but the calculation time increases significantly with larger ones. On the other hand, while SPU's calculation time is slower for small matrices, it increases only marginal with matrix size, ultimately making it faster and more reliable.
\begin{table}[h]
    \centering
    \caption{\centering Calculation time comparison in seconds for matmul operations with Concrete-ML and SPU execution time as t$_{SPU}$ and t$_{ConcrML}$.}
    \begin{tabular}{ c|c|c } 
        matrix size ($m \times n$)& t$_{SPU}$ & t$_{Concr-ml}$ \\
        \hline
        $10\times10$ & 0.05896 & 0.0275 \\ 
        $50\times50$ & 0.06344 & 0.6129 \\
        $100\times100$ & 0.07816 & 2.7886 \\
    \end{tabular}
    \label{tab:matmul-smpc-fhe-time}
\end{table}

\paragraph{Activation functions.}
The results for the activation functions are shown in Table~\ref{tab:activations-smpc-fhe-time}. The SPU computation time stays constant with respect to vector size, while rising slightly with activation function type. Concrete-ML, on the other hand, exhibits a steady increase with regards to the vector size, but remains relatively constant with different activation functions. Nevertheless, SPU is much faster.
\begin{table}[h]
    \centering
        \caption{\centering Calculation time comparison in seconds for activation functions with Concrete-ML and SPU execution time as t$_{SPU}$ and t$_{Concr-ml}$.}
    \begin{tabular}{ c|c|c|c } 
        function & vec. size & t$_{SPU}$ & t$_{ConcrML}$ \\
        \hline
        ReLU & 8 & 0.04829 & 2.5413 \\ 
        & 16 & 0.04813 & 4.6703 \\
        & 32 & 0.04822 & 8.1009 \\ \hline
        GELU & 8 & 0.06403 & 2.5463 \\ 
        & 16 & 0.06427 & 4.6865 \\
        & 32 & 0.06450 & 8.1087 \\ \hline
        Sigmoid & 8 & 0.07755 & 2.5293 \\ 
        & 16 & 0.07732 & 4.6631 \\
        & 32 & 0.07805 & 8.0897 \\
    \end{tabular}
    \label{tab:activations-smpc-fhe-time}
\end{table}

\subsection{Regressions}

This section compares the inference times of FHE and SMPC across various regression models.
Two models were benchmarked: linear and random forest regression.
We tested the inference time for different numbers of features and for the random forest regressor, we also varied the number of estimators.
The Scikit-learn implementations of Concrete-ML were used as a basis for measuring plaintext inference times.
For FHE encrypted inference the same was used over encrypted data. Furthermore the highest allowed quantization of eight bits was applied.
This allows for the best accuracy, but the slowest performance. Thus speeding up FHE is possible.
For SMPC both regressions were implemented using the JAX library as a single-layered linear network, without activation for linear regression and with sigmoid activation for logistic regression.
The results were calculated using fixed-point 64-bit numbers.

\paragraph{Linear Reg.}
Linear regression is the most basic regression.
The results of the comparison are shown in Table \ref{tab:linear-regression-comparison}.
SPU inference times appear to remain constant at approximately 0.079 seconds regardless of the number of features.
Compared to plaintext inference, however it is on average slower by a factor of approximately 600.
Fully-homomorphic encryption follows a linear increase in inference time with respect to the number of features. It is slower than plaintext, but faster then SMPC for all tests.
This is especially true for a small number of features, where Concrete-ML is up to 17 times faster.  

\begin{table}[h]
    \centering
    \caption{\centering Evaluation time comparison in seconds for linear regression with Concrete-ML and SPU execution time as t$_{SPU}$ and t$_{ConcrML}$.}
    \begin{tabular}{c|c|c|c}
        features & t$_{plain}$ & t$_{SPU}$ & t$_{Concr-ml}$ \\ \hline
        10 & 15e-05 & 0.0791 & 0.0046\\
        50 & 7e-05 & 0.0790 & 0.0095\\
        100 & 23e-05 & 0.0786 & 0.0158\\
        250 & 11e-05 & 0.0793 & 0.0451\\
        500 & 18e-05 & 0.0794 & 0.0777\\
    \end{tabular}
    \label{tab:linear-regression-comparison}
\end{table}

\begin{figure}[h]
    \centering
    \includegraphics[width=\linewidth]{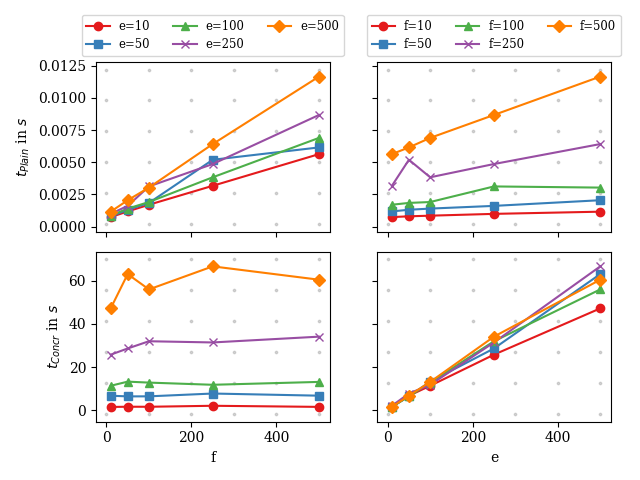}
    \caption{Random Forest Regressor benchmarks.}
    \label{fig:forest_benchmark}
\end{figure}

\paragraph{Random Forest.}
For the last regression, we decided to use the random forest regression.
It remains a popular model, particularly in fields such as banking, trading, medicine, and e-commerce \cite{random_forest}.
The results of our tests in FHE are presented in Figure \ref{fig:forest_benchmark}.
The first row portrays the inference times for plaintext and the second shows the times for encryption messages.
The first column presents the inference time depending on the number of features and the second in regards to the estimators.
Seemingly, the graph's behaviours are flipped.
Plaintext inference increases linearly with more features and remains constant for different estimators.
Conversely, FHE-encrypted inference grows linearly with the number of estimators and remains constant with different amounts of features.
The function with 500 estimators in the lower left graph behaves differently, decreasing first before becoming stable. 
One possible reason is, that the compilation to an FHE circuit for that configuration is not efficient.
We can explain the seemingly flipped behaviour of the plain and encrypted inference times by examining FHE and the structure of the random forest regressor.
In a random forest regressor a number of trees, or estimators, are created over a sample of features. For inference, all of them are traversed and a mean is returned.
To traverse these trees many comparisons and some simple additions are performed. In FHE these comparisons are very expensive.
Thus, the more trees that exist, the longer it takes to evaluate them all.
This is not the case for plaintext prediction, where the aggregation of the trees for parameter selection may create more overhead with more features.
Overall we can see, that encrypted evaluation takes between five and more than 20 seconds. Therefore, it is not feasible to repeatedly calculate these on consumer-level computers without GPU. Evaluations on servers, with good GPUs could however be applicable. 
For SPU the Random Forest is not natively implemented in a JAX library, so it was omitted due to time constraints.

\subsection{Deep Learning}

In this section multiple basic deep learning models are benchmarked.
The focus was put on the widely used and generic structures.
These include Dense and Convolutional models. 
Variations of different depths and widths were tested.
For the FHE benchmarks the Concrete-ML compilation uses six bit quantization, which yields lower accuracy but better performance.
We use this because the compiler failed to convert the models into FHE circuits with higher quantization.
Reasons for this are, that certain calculations would have exceeded the 16-bit threshold for table lookups, or that it wasn't able to find an allowed set of parameters to adhere to security standards. 

\begin{figure}[h]
    \centering
    \includegraphics[width=\linewidth]{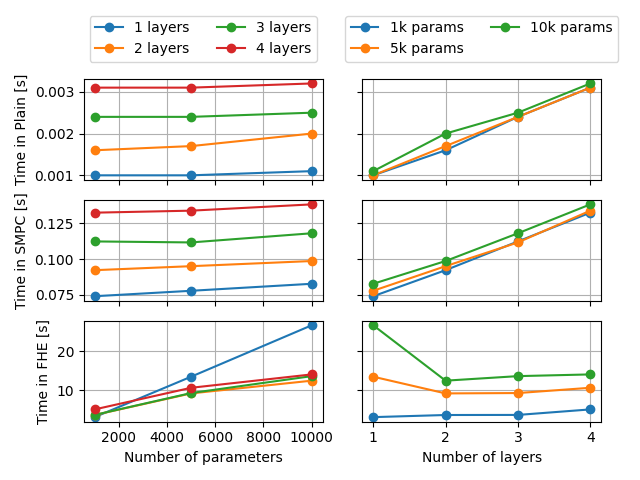}
    \caption{Dense Model benchmarks.}
    \label{fig:dense_model_bench}
\end{figure}

\paragraph{Dense Models.}
Dense models consist of only linear layers and activation functions between the layers~(ReLU in this case).
This type of model is sometimes referred to as Multi-Layer Perceptron~(MLP).
We tested a range of trainable parameters~(1000--10000 parameters) and a range of layers~(1--4 layers).
The first row of Figure \ref{fig:dense_model_bench} portraits the results of plaintext inference.
On the left side the inference time is shown in relation to the number of parameters and the second column illustrates the evaluation time depending on the amount of layers.
We observe the default linear behaviour: as the number of layers increases, so does the inference time.
Conversely, when the number of parameters increases, the inference time remains constant with respect to the number of layers.
The results for SMPC are shown in the second row of Figure~\ref{fig:dense_model_bench}, with the same column alignment as above.
As before, the same relationship regarding both parameters and number of layers can be observed.
The inference time increases, but the relationships remain consistent.
The results of the FHE evaluation are shown in the last row.
It can be observed, that the relationship is seemingly flipped. 
The number of layers does not appear to significantly impact the inference time. 
On the other hand, the number of parameters seems to linearly impact all the tested models, with the one-layer network having the greatest effect on the inference time. 
Additionally, the inference time for models with one layer is significantly higher then for more layers in the 5k and 10k parameter configuration.
One possible reason is, that the individual calculation times in FHE for each parameter are much more time consuming, than the potential speedup from parallelization by having more parameters in one layer.
This is reinforced by the fact, that we only ran inference on a 15 core CPU, which severely limited the amount of parallel execution for 1k nodes.

\begin{figure}[h]
    \centering
    \includegraphics[width=\linewidth]{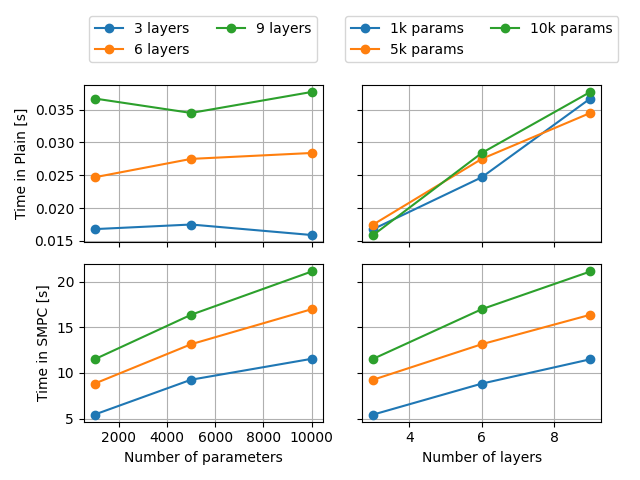}
    \caption{Plain and SMPC CNN benchmarks.}
    \label{fig:cnn_benchmark}
\end{figure}

\paragraph{Convolutional Models.}
Convolutional neural networks~(CNNs) are a family of networks primarily used for image processing. Their main building blocks are convolutional layers. For benchmarking, a simple model was constructed, that works with $3\times112\times112$ randomly generated pictures. A single layer of a generic CNN consists of a convolution operation, maximum pooling and a ReLU activation function. Once again the benchmark tries variations in depth~(3,6,9 layers) and number of parameters~(1000--10000). The plaintext results are shown in the first row of Figure~\ref{fig:cnn_benchmark}. Similarly to MLP for plaintext inference, the relationship between the number of layers and inference time is roughly linear, while the relationship between the number of parameters and inference time remains fairly constant. The results for SMPC can be seen in the second row of Figure~\ref{fig:cnn_benchmark}. Analogous behaviour to the plaintext inference can be observed when changing the number of layers. However, the relationship between the number of parameters and the inference time (column one) is now linear, which differs from the plaintext implementation. 

\begin{figure}[h]
    \centering
    \includegraphics[width=\linewidth]{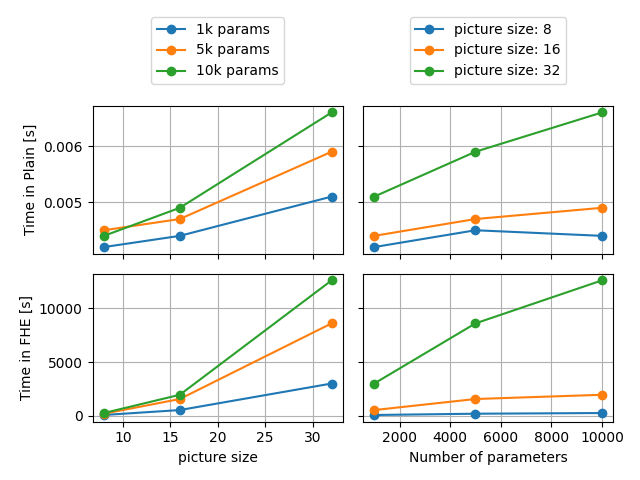}
    \caption{Plain and FHE CNN benchmarks.}
    \label{fig:cnn_fhe_benchmark}
\end{figure}

Benchmarking FHE inference times for RGB images of size $112\times112$ takes too long, to yield any reasonable results, so we evaluated smaller inputs and varied the number of parameters in the same model. We tested RGB images of size $8\times8$, $16\times16$ and $32\times32$. For these small inputs, deeper networks could not be benchmarked reliably because activations collapsed to zero, producing non-representative runtimes. The results are shown in Figure \ref{fig:cnn_fhe_benchmark}. The left column shows the evaluation time depending on the picture size and the right column in regards to the parameters for plaintext in the first row and for encrypted text in the second. As we can see, ciphertext and plain inference behave very similar. With larger images, the evaluation time increases roughly quadratically for both instances. Changing the number of parameters results in a sharp increase in calculation time, roughly doubling or tripling when going from 1k to 5k parameters, and increasing by about 50\% going to 10k. This suggests a square root relationship between the number of parameters and the inference time. Furthermore inference times, even for those small picture sizes, can range up to over 3 hours, making them unusable in practice.  

\section{\uppercase{Discussion / Conclusions}}%
\label{sec:discussion_conclusion}

%\paragraph{Summary recommendations.}
Our results show, that both secure computation technologies have their strengths and weaknesses.
SMPC performs very well with larger models.
It can handle dense models and convolutional neural networks with no problem, making it applicable for localhost settings in real world scenarios.
As discussed in the protocol comparison, SMPC is heavily affected by network artifacts, such as packet loss and transmission times.
These factors can  significantly impact inference times. Nevertheless, we expect SMPC to perform very well still.
We advise using close proximity networks or very fast channels.
Regression models and simple calculations are also very fast, taking only up to 100ms to compute.
For practitioners in need of a secure computational technology, SMPC can be used for all use cases.
However, it is difficult to setup because one needs multiple servers.

This is not the case for FHE, which is really easy to deploy and performs better in linear regressions.
Furthermore, Concrete-ML offers a wide variety of pre-implemented regression and classifier models, making their deployment very easy.
It additionally is possible to implement so called hybrid models, which split inference between user (in clear) and server (encrypted).
Those may speed up inference time for models considerably.
Additionally, FHE can achieve a significant performance boost by using GPUs or other FHE based software and hardware.
Such a speed-up is probably much smaller in SMPC because overhead from network artifacts consume most of the time.
Larger models, such as CNNs, LSTMs or Transformers, are either too inefficient or not supported in FHE.
Thus we recommend using it for simple models and calculations, especially with GPUs, as it is much easier to set up and faster than SMPC. 

\paragraph{Conclusions.}
In summary, FHE and SMPC are well-researched technologies that are often studied independently.
However, our direct comparison provides a clearer picture of both technologies and helps practitioners to understand them, which in turn fosters adoption and research.
We present strengths and weaknesses of each technology based on our benchmarking results, which can be vital for choosing the right technology for specific applications.
Although both technologies can perform all computations, our results show that significant trade-offs must be made when choosing one over the other for a specific application, e.g. CNN.
Nevertheless, this is only the first step, and more research is needed on the standardised benchmarking of MPC and FHE in machine learning.

\paragraph{Future work.}
Because FHE and SMPC are both very dynamic fields, the impact of new protocols and advancements should be monitored and studied further.
Additionally, we only performed the benchmarks using general purpose CPUs and the analysis of the impact of GPUs or FHE accelerators on the performance will be essential for a complete picture.
Future research should also focus on the tailoring and optimization of models for FHE or SMPC, e.g., by minimizing the use of non-linear functions.

% - Grafikkarten: FHE boost MPC weniger weiss man nicht genau
% - FHE acceleratoren
% - Impacts of new protocols, because dynamic field
% - Crypto friendly ML designs and combined simulators and compilers who automatically help in performance estimation ...

\section*{\uppercase{Acknowledgements}}
%If any, should be placed before the references section without numbering. To do so please use the following command: 
%\textit{$\backslash$section*\{ACKNOWLEDGEMENTS\}}
This work was in part funded by the European Union under the HORIZON SESAR JU Grant Agreement No. 101114675 (HARMONIC), and the Austrian Research Promotion Agency FFG within the PRESENT project (grant no. 899544). 
%Views and opinions expressed are however those of the author(s) only and do not necessarily reflect those of the funding agencies.
%Neither the European Union, FFG, nor the granting authority can be held responsible for them.

\bibliographystyle{apalike}
{\small
\bibliography{references/references.bib}}

%\section*{\uppercase{Appendix}}
%If any, the appendix should appear directly after the references without numbering, and not on a new page. To do so please use the following command: \textit{$\backslash$section*\{APPENDIX\}}

\end{document}